\documentclass[10pt,a4paper,preprint,nofootinbib,floatfix,number]{elsarticle}

\usepackage{subcaption}
\usepackage{enumitem}
\usepackage{hyperref}

\begin{document}

%Coevolution in the model of social interactions on scale-free networks
\title{Coevolving complex networks in the model of social interactions}
\date{\today}

\author[past]{Tomasz Raducha}
\ead{tomasz.raducha@fuw.edu.pl}
\author[past]{Tomasz Gubiec}
\address[past]{Institute of Experimental Physics, Faculty of Physics, University of Warsaw, Pasteura 5, 02-093 Warsaw, Poland}

\begin{abstract}
We analyze Axelrod's model of social interactions on coevolving complex networks. We introduce four extensions with different mechanisms of edge rewiring. The models are intended to catch two kinds of interactions - preferential attachment, which can be observed in scientists or actors collaborations, and local rewiring, which can be observed in friendship formation in everyday relations. Numerical simulations show that proposed dynamics can lead to the power-law distribution of nodes' degree and high value of the clustering coefficient, while still retaining the small-world effect in three models. All models are characterized by two phase transitions of a different nature. In case of local rewiring we obtain order-disorder discontinuous phase transition even in the thermodynamic limit, while in case of long-distance switching discontinuity disappears in the thermodynamic limit, leaving one continuous phase transition. In addition, we discover a new and universal characteristic of the second transition point - an abrupt increase of the clustering coefficient, due to formation of many small complete subgraphs inside the network.
\end{abstract}


\maketitle

%%%%%%%%%%%%%%%%%%%%%%%%%%%%%%%%%
%                                                                                       %
%                              INTRODUCTION                                   %
%                                                                                       %
%%%%%%%%%%%%%%%%%%%%%%%%%%%%%%%%%

\section*{Copyright}
\noindent \textcopyright 2016. This manuscript version is made available under the CC-BY-NC-ND 4.0 license
\url{http://creativecommons.org/licenses/by-nc-nd/4.0/}
\\

\noindent You can find the original article at \url{http://dx.doi.org/10.1016/j.physa.2016.12.079}

\section{Introduction}\label{section:int}

Numerous of recently observed phenomena can be described by network models. In many cases it is a natural choice, because in the structure of the system one can clearly distinguish vertices and edges between them. This approach to describe the dynamics of various systems is applied in a wide range of problems \cite{dorogovtsev2002evolution,albert2002statistical}. The complex structures studied by physicists, biologists, economists and sociologists, as well as phenomena occurring on the border of these sciences can be captured using graph theory \cite{kim2008complex}. The flexibility is so high thanks to an arbitrary interpretation of vertices and connections. In physics it may be atoms, in biology neurons, in economics financial institutions and in sociology people or social groups. Assuming two-body interactions, the limitation is the discreet nature of the phenomenon, but only in the structure, because the connections can be weighted to indicate for example the distance between cities, or the strength of the interaction, and vertices can be described by any state determining for instance the energy or opinion. In recent times, a great deal of effort has been devoted to analyzing social networks, examining the dynamics of opinion or culture \cite{nyczka2012phase,nyczka2013anticonformity,axelrod1997dissemination,klemm2003,gandica2013thermodynamic,genzor2015thermodynamic,laguna2003vector}, the spread of epidemic, information or innovation \cite{pastor2001epidemic,pinto2016setting,przybyla2014diffusion} etc. In this paper we consider the model of social interactions also referred to as the model of dissemination of culture \cite{axelrod1997dissemination}, in which every node represents one individual having its own preferences, interests, attitude, opinions, religion, hobbies and so on, collected under the general term \emph{culture}. Culture of each individual is represented by a vector of natural numbers, each number being one \emph{trait} of that person. Connection between nodes indicates that two individuals can interact and receive traits from each other, meaning they can be neighbors, co-workers, friends or family in the real world.

The model of dissemination of culture was originally defined by Robert Axelrod on a static square lattice. He identified basic dependencies of the homogeneity of the system on four parameters - the number of traits $F$, the number of possible values of each trait $q$, the size of the network $N$ and the size of the neighborhood (4, 8 or 12 neighbors on a square lattice). Since his publication, the model has been further investigated by other authors. First of the most important results is the discovery of the order-disorder non equilibrium phase transition in parameter $q$ \cite{castellano2000nonequilibrium}. In the very same work authors show that system's behavior for $F = 2$ traits differs significantly from $F > 2$. In the first case the phase transition is continuous, while in all other cases the transition is discontinuous. The size distribution of the domains is different as well, but universal for all $F > 2$. The suggestion of Axelrod to apply cultural drift was also analyzed in the form of a random noise in values of traits \cite{klemm2003global,kim2011effects}, which resulted in no phase transition in $q$ in this scenario. However, a new phase transition in effective noise rate occurred. Important results have been given by Maxi San Miguel et. al. \cite{klemm2003nonequilibrium}, who studied the model of social interaction on complex networks. The outcome of this casework indicates that the structure of the graph may and does influence the appearance of a phase transition. In particular, a scale-free network with high clustering coefficient displays a phase transition even in the thermodynamic limit. There were more interesting approaches: injecting intolerance \cite{gracia2011coevolutionary,gracia2011selective}, adding acceptance or discussion in the interaction of two vertices \cite{dybiec2012accepting}, applying conservativeness or nonconformity \cite{dybiec2012conservativeness,parravano2007intracultural}, and taking into account physical and social distance between agents \cite{pfau2013co,stivala2014ultrametric}, to mention only a few.

The majority of the proposed models consider a static graph, representing a group of people, with dynamic states of vertices. However, social networks are not static and their structure may change over time. Furthermore, in longer time scales the topology may have a crucial impact on the behavior of the system. If we want to analyze the spread of the culture, we should leave the static picture of the network aside. It is hard to imagine that in time scales corresponding to cultural transformations the network of contacts remains static. Maxi San Miguel et. al. \cite{sanmiguel2004coev,sanmiguel2007} - the first, who took this fact into account - analyzed the model of social interactions on coevolving networks. This was an important step, approaching the original Axelrod's model to real-world social networks. However, in the words of the author ,,\textit{The results do not depend on the initial network topology, because the repeated rewiring dynamics leads to a random network with a Poisson degree distribution}''. Indeed, in the model of coevolution all the networks generated during the simulation have a random structure. But real-world social networks are as distant from random networks as they are from regular ones \cite{watts1998collective,albert2002statistical} and the structure of the network may have a vast influence on its functionality \cite{albert2000error,cohen2000resilience}. Therefore, in this paper we analyze the coevolution in Axelrod's model, taking into account different mechanisms of switching edges to imitate the important features of social networks, i.~e. high value of the clustering coefficient and the power-law degree distribution.

%%%%%%%%%%%%%%%%%%%%%%%%%%%%%%%%%
%                                                                                       %
%                                   MODEL                                         %
%                                                                                       %
%%%%%%%%%%%%%%%%%%%%%%%%%%%%%%%%%

\section{Model}\label{section:model}

We start with a random graph with $N$ vertices each representing one agent. Every node $i$ is described by a vector $\sigma_i = (\sigma_{i, 1}, \sigma_{i, 2}, ..., \sigma_{i, F})$ ($F$ features of an agent). Every feature can initially adopt one of $q$ discrete values $\sigma_{i, f} \in \{1, 2, ..., q\}, f = 1, 2, ..., F$, which gives $q^F$ possible different states. To begin, we draw a set of $F$ traits for each node with equal probability for every value form $1$ to $q$. Next:

\begin{enumerate}
\item Draw active node $i$ and one of its neighbors\footnote{In case of coevolving networks "neighbors" are just pairs of connected vertices.} $j$.
\item Compare vectors $\sigma$ of chosen vertices - determine the number $m$ of identical traits (overlap), such that $\sigma_{i, f} = \sigma_{j, f}$.
    \begin{itemize}
    \item If all traits are equal i. e. $m = F$, nothing happens.
    \item If none of the traits are equal $m = 0$, disconnect the edge $(i, j)$ from node $j$, draw a new node\footnote{Methods of drawing a new neighbor can be different.} $l$, and attach a link to it, creating an edge $(i, l)$.
    \item In other cases, with probability equal $m/F$ an interaction occurs, in which we randomly select one of the not-shared features~$f'$ (from among $F - m$) and node $i$ adopts its value from node $j$, i. e. $\sigma_{i, f'} \to  \sigma_{j, f'}$.
    \end{itemize}
\item Go to the next time step.
\end{enumerate}

The outlined algorithm differs from Axelrod's model in the situation of no overlap $m = 0$. Originally, nothing happened in such case, while in our model a rewiring of the connection occurs, thus implying a coevolution of agents' states and the structure of the graph itself. This form of the model was at first presented by San Miguel et. al. \cite{sanmiguel2007}. The main difference and also the clue of our model is the method of drawing new neighbors for the active node in case of no overlap. The in-depth analysis of the model with uniform probability distribution for every node in switching event showed that it leads to Poisson degree distribution regardless of the initial conditions. To achieve a complex structure of graphs generated during the simulation we propose four modifications of the model of social interactions with coevolution. Every modification introduces an alternative way of selecting a new neighbor (from among all nodes) in rewiring event. The probability of drawing node $i$ with degree $k_i$ is as follows:

\begin{enumerate}
\item Model A with preference of high degree:
\begin{itemize}
\item $P_A(i)~=~\frac{k_i}{\sum_{j=1}^{N} k_j}$.
\end{itemize}
\item Model B with preference of high degree with auto-connections:
\begin{itemize}
\item $P_B(i) = \frac{k_i + 1}{\sum_{j=1}^{N} k_j + 1} $.
\end{itemize}
\item Model C  with square preference of high degree:
\begin{itemize}
\item $P_C(i)~=~\frac{( k_i + 1 )^2}{\sum_{j=1}^{N} ( k_j + 1 )^2}$.
\end{itemize}
\item Model D with preference of close neighborhood:
\begin{itemize}
\item switching to nodes distant by two edges (to neighbors of neighbors).
\end{itemize}
\end{enumerate}

The first model (A) uses the well-known Barab\'{a}si-Albert mechanism of preferential attachment \cite{barabasi1999emergence}. The second model (B) has a small modification - it may be regarded as counting virtual auto-connections (not real auto-connections, which are prohibited), so there is always a nonzero probability for every node to be selected. In the model A, if a node loses its last connection, it remains lonely to the end of the simulation. The third model (C) also counts virtual auto-connections, but the probability increases with the square of the degree, which should in theory lead to a formation of big hubs in the network.
We decided to analyze such a model, because in coevolving models of social interactions the nodes can lose their connections. Moreover, the probability of losing a connection grows with the degree of the node. By contrast, in the original Barab\'{a}si-Albert algorithm (of generating scale-free networks) once an edge is assigned to a vertex it can never be erased. Hence, the model C is an attempt to compensate for the loss of connections by higher probability of forming new ones. In all models auto-connections and double (multiple in general) connections are not allowed. Therefore, in the first three models (A, B, C), during the rewiring procedure, if the active node itself or any of its neighbors are chosen as the new neighbor, the procedure is repeated. The fourth model (D) is essentially different. The new neighbor is drawn from the uniform distribution on the set of neighbors of neighbors of the active node, excluding the active node itself and its neighbors, i.~e. the set of nodes distant by two edges. This set is much smaller than in other models. In consequence, this rewiring mechanism cannot generate long distant connections, but it provides local growth of the clustering coefficient at every switch by definition \cite{klemm2002highly}.

It is worth mentioning that for $q \to \infty$, when no two nodes have any common traits and there are no interactions, only switching, we are able to analytically derive the probability distribution for choosing new neighbors, which will lead to the power-law distribution of the nodes' degree \cite{dorogovtsev2003principles}. In particular, the second model (B) for $q \to \infty$ should result in the exponential distribution of the degree. In section \ref{section:stucture} we can see that this result is also a good approximation for finite values of $q$, e.g. for $q = 5000$ in the model B during the simulation only in $75\%$ of time steps rewiring occurs, but the degree distribution is still exponential (see. FIG.~\ref{fig:degree}). However, the most interesting behavior and true coevolution occurs for smaller values of $q$, where interactions start playing an important role. Yet, taking the traits of nodes into account renders the analytical solution non-trivial. We, therefore, decided to investigate all models numerically.

The first three models (A, B, C) favor, to a different extent, the nodes with a high degree, which means that the hubs are more likely to obtain new connections. It is well known that this behavior is typical for networks of citations in science and a collaboration of scientists or actors \cite{newman2001clustering,jeong2003measuring}, but is this the real microscopic view of everyday interactions in society? One of the possibilities is that we look for new friends through existing relations, without relaying on someone's popularity. The model D was created to capture this observation.

\begin{figure}
\captionsetup[subfigure]{labelformat=empty}
\begin{subfigure}{.5\textwidth}
  \centering
  \includegraphics[width=1.0\linewidth]{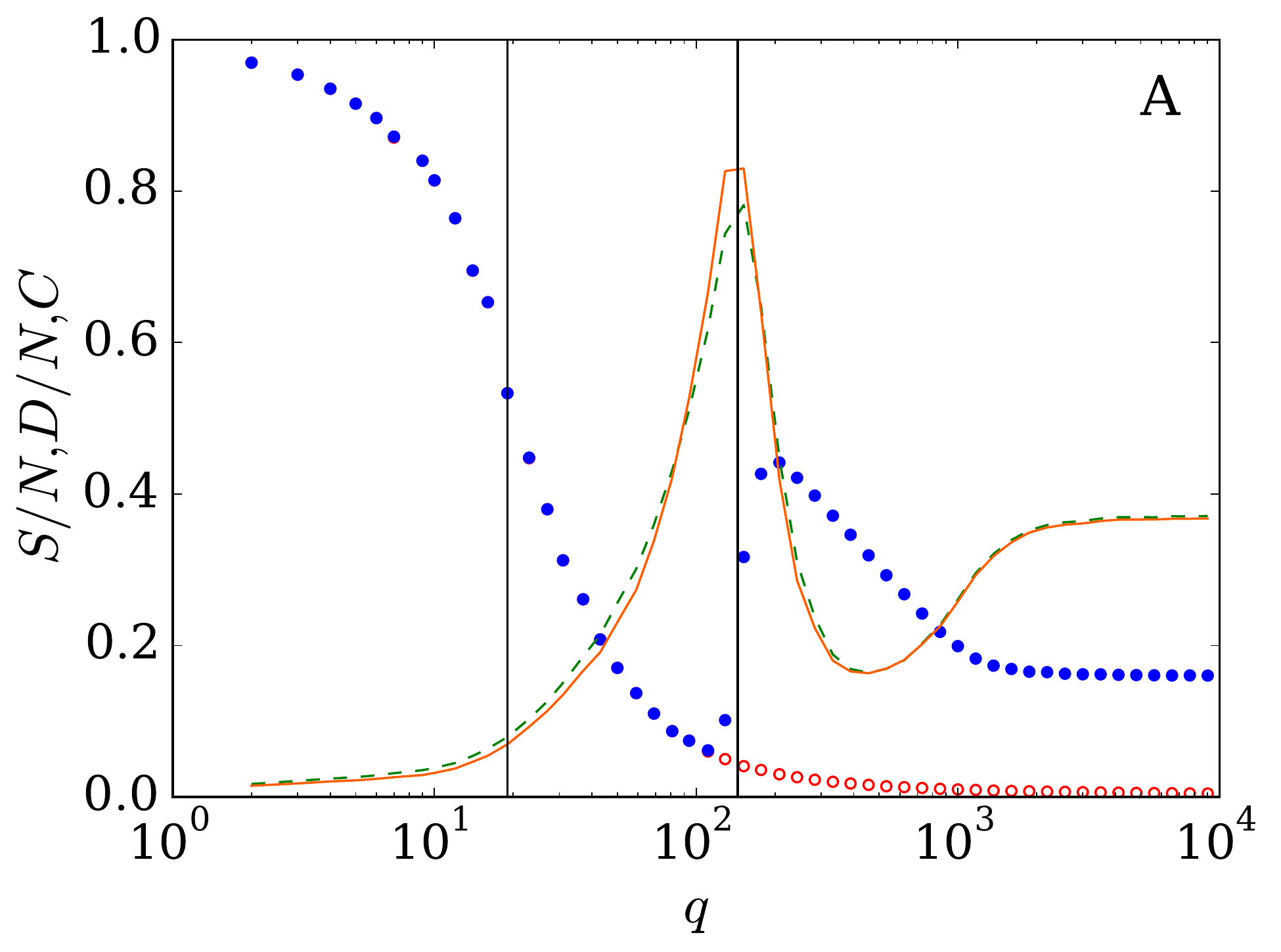}
  \caption{\label{sub:BA}}
\end{subfigure}
\begin{subfigure}{.5\textwidth}
  \centering
  \includegraphics[width=1.0\linewidth]{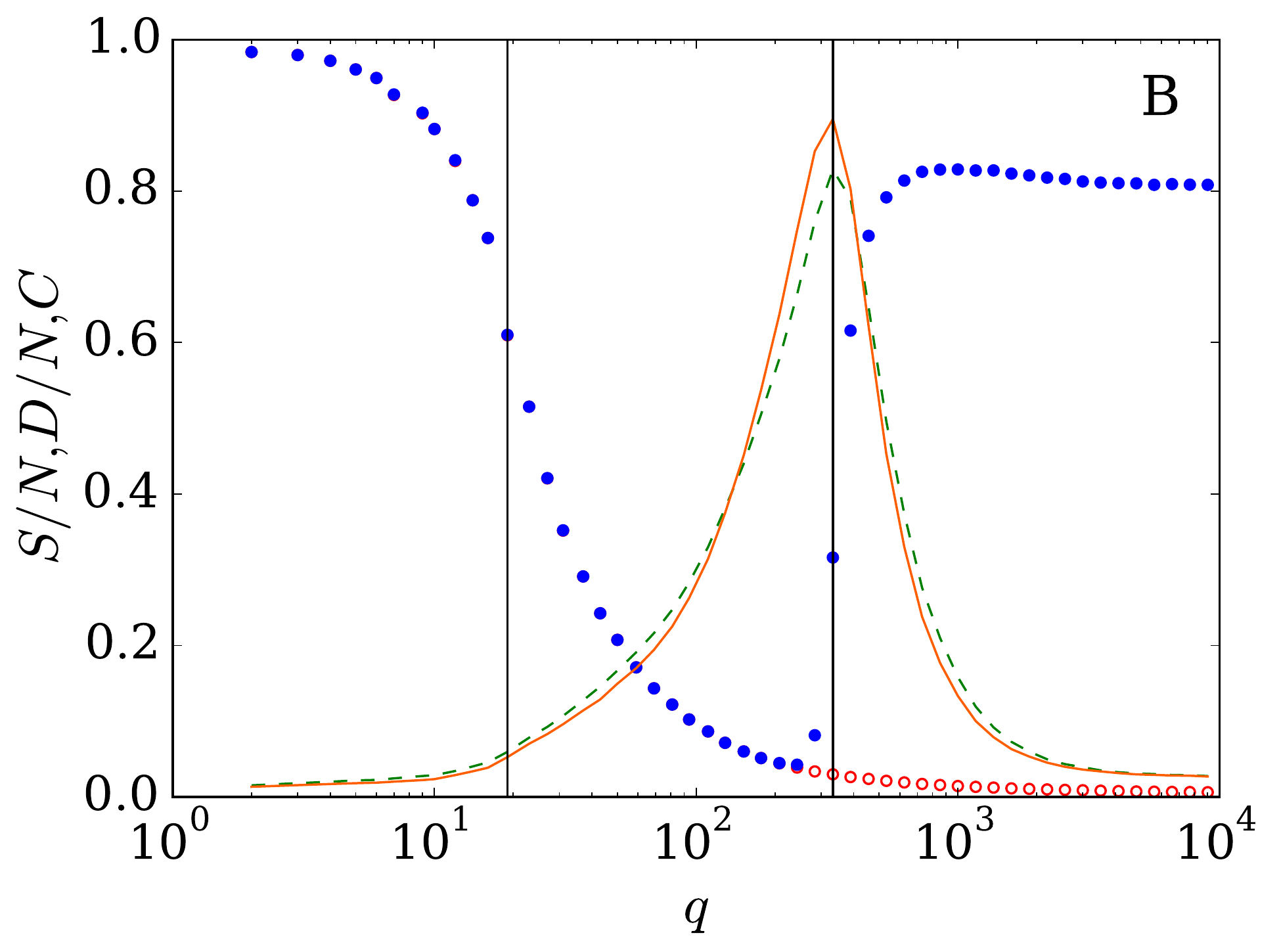}
  \caption{\label{sub:k1}}
\end{subfigure}
\begin{subfigure}{.5\textwidth}
  \centering
  \includegraphics[width=1.0\linewidth]{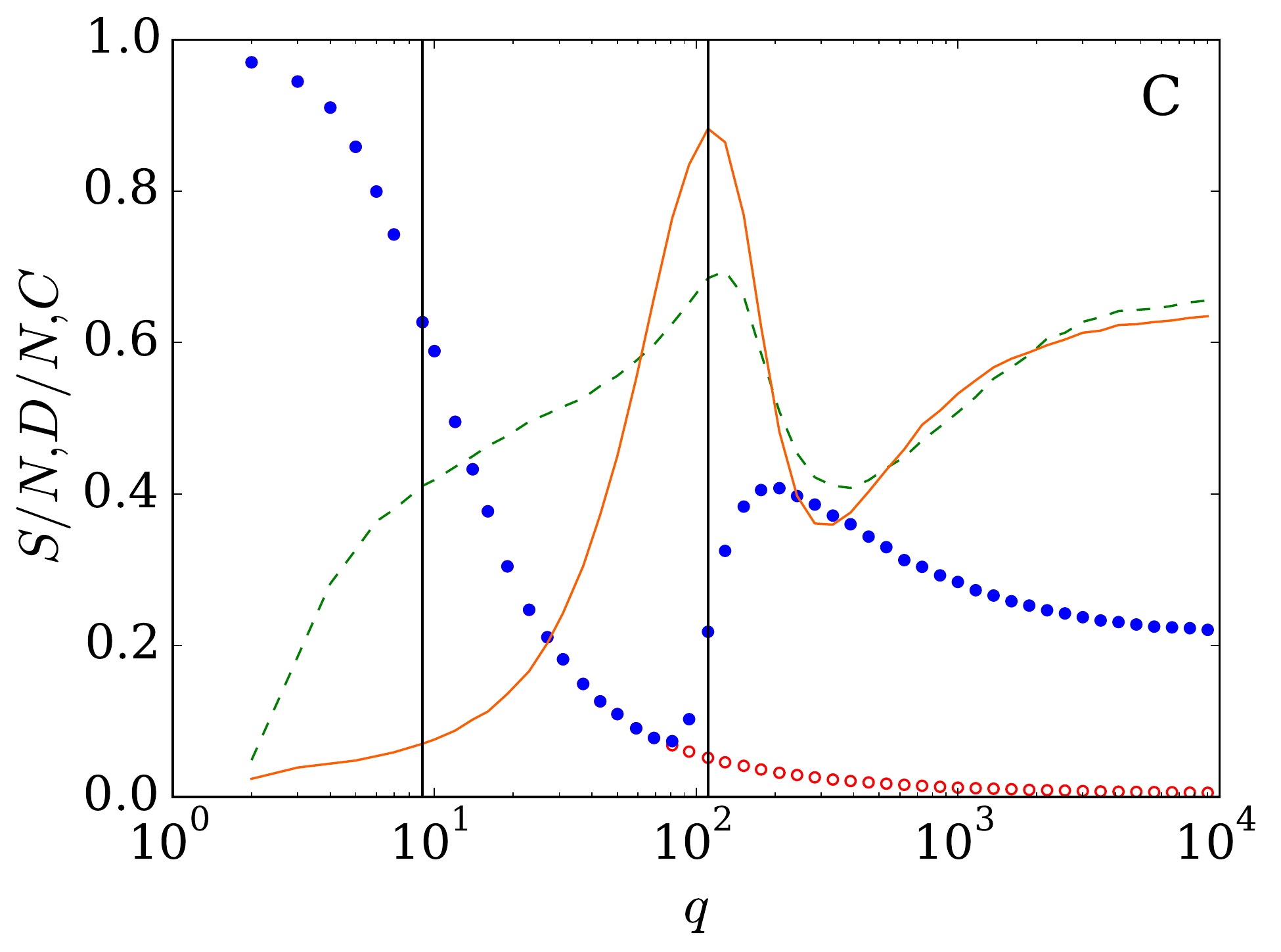}
  \caption{\label{sub:k2}}
\end{subfigure}
\begin{subfigure}{.5\textwidth}
  \centering
  \includegraphics[width=1.0\linewidth]{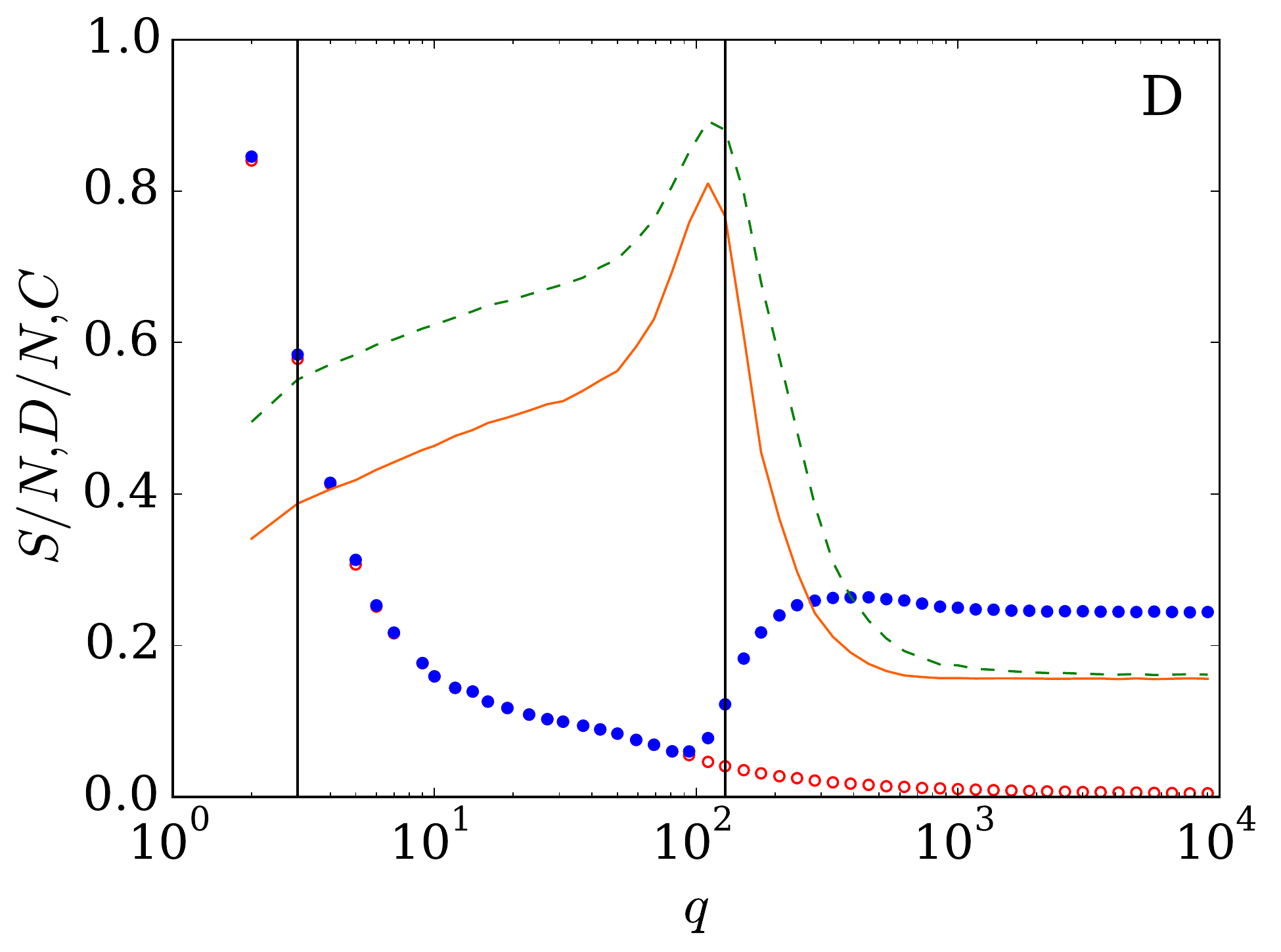}
  \caption{\label{sub:clust}}
\end{subfigure}
\begin{subfigure}{.5\textwidth}
  \centering
  \includegraphics[width=1.0\linewidth]{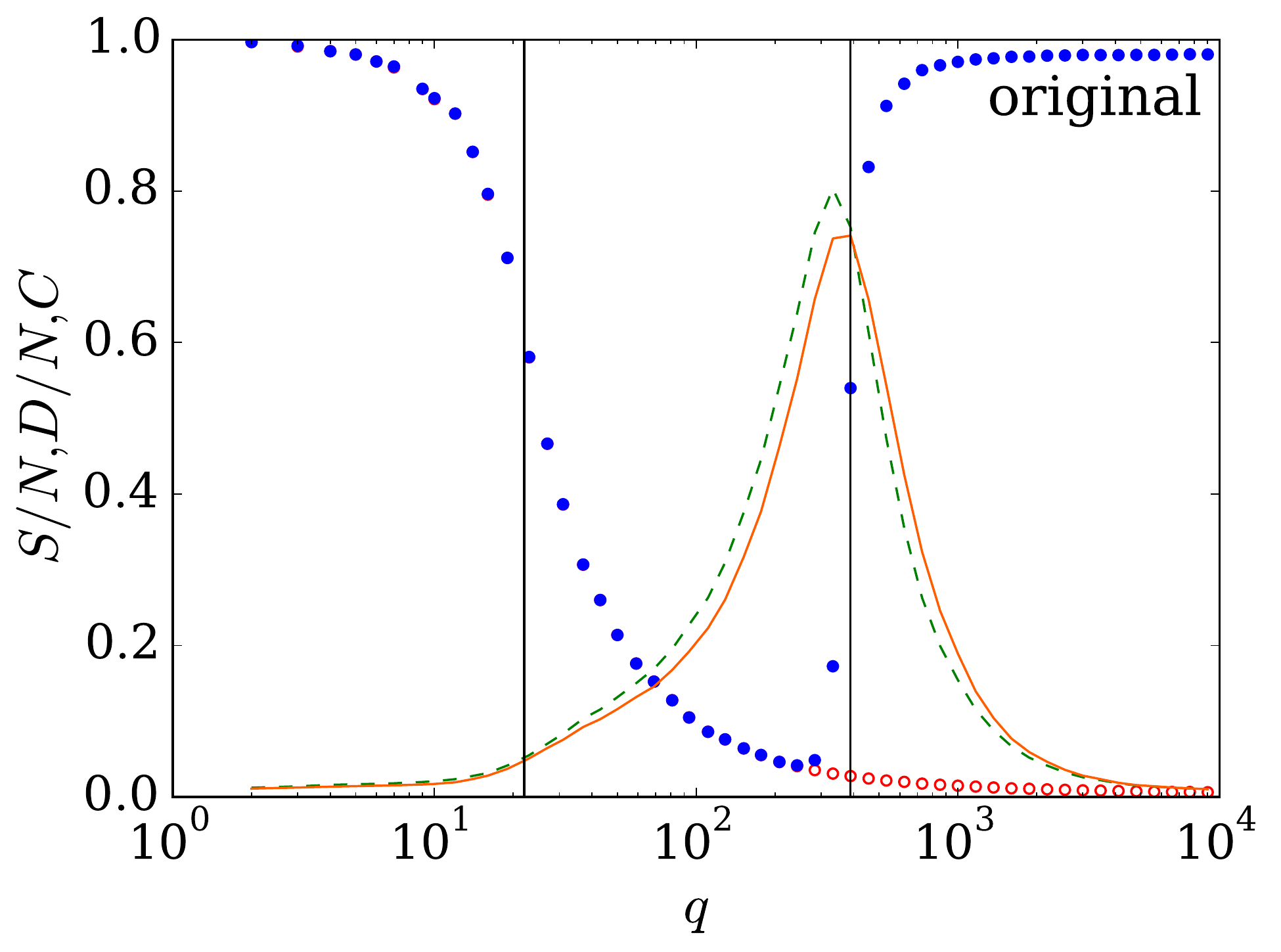}
  \caption{\label{sub:normal}}
\end{subfigure}
\caption{Average relative size of the largest network component (full circles) and largest domain (empty circles),
global (solid line) and average local (dashed line) clustering coefficient in the stationary configuration vs $q$, for $N = 500$,
averaged over 400 realizations. Character in the upper right corner represents the number of a model.
Vertical lines are placed at $q=q_c$ and $q=q*$, which are respectively 19 and 144 (model A),
19 and 333 (model B), 9 and 111 (model C), 3 and 129 (model D), 22 and 389 (original model).\label{fig:phases}}
\end{figure}

%%%%%%%%%%%%%%%%%%%%%%%%%%%%%%%%%
%                                                                                       %
%                              PHASE DIAGRAM                                %
%                                                                                       %
%%%%%%%%%%%%%%%%%%%%%%%%%%%%%%%%%

\section{Phase diagram}\label{section:phasediagram}

We performed extensive numerical computations for all models from the previous section, including the model proposed by San Miguel et. al. \cite{sanmiguel2007} as a reference case. All simulations were performed for the number of traits $F = 3$, since it was shown that for $F = 2$ the model has a significantly different behavior than for $F > 2$ \cite{castellano2000nonequilibrium}. We also decided to analyze graphs with the average degree $\langle k\rangle = 4$, in order to keep the nodes' neighborhood (on average) close to the square lattice. To see the influence of the new switching rules on network's topology, not obscured by the initial state, we begin every simulation from a random graph \cite{ER1959random,ER1960evolution}, at first with $N = 500$ agents.

In all of the models we obtained three phases (FIG.~\ref{fig:phases}). To describe them, we introduce two order parameters. The first one is the largest component $S$, where by component $s$ we mean a connected subgraph of the network, which is not a subgraph of another connected subgraph of the network. The second parameter is the largest domain $D$, and domain $d$ is a connected subgraph with one and the same set of traits for every node. By definition, a given domain cannot exceed the size of the component it shares the nodes with. On the other hand,  the number of components cannot be superior to the number of domains.

The phase I is generally characterized by a single big component of size comparable to the size of the network, and a few small components. The size of the largest domain is equal to the size of the largest component. By increasing $q$ we obtain an order-disorder phase transition at point $q_c$. In the phase II the network disintegrates into multiple small components, each containing exactly one domain. The first two phases are static - the simulation always ends in a gridlock configuration, where no further rewiring nor interaction can occur. It is worth mentioning that the state of the network is frozen, if and only if every component contains exactly one domain. Further increase of $q$ leads to a second phase transition at point $q^*$, in which a recombination occurs. In the phase III the largest component is bigger than in phase II, but the size of the largest domain still decreases in $q$. In this phase the final state is dynamic - the connections are constantly being rewired and the number of components is fluctuating around an average value. The level of recombination in the phase III is different in all models - $\langle S\rangle$ varies from $0.16$ in the model A to $0.8$ in the model B for $q \approx 10^4$. For $q > 10^3$, in all models, the network contains one big component  and multiple tiny components containing at most just several nodes, but most often only a single one. The isolation of single nodes is natural in the model A - once a node loses all its links there is no possibility to chose it for a neighbor. The second model (B) fixes this behavior, as it gives a nonzero probability for lonely nodes, thus it looks similar to the model of San Miguel et. al. \cite{sanmiguel2007} in terms of the phase diagram, but the graph does not fully recombine, since $\langle S\rangle$ stops at $0.8 N$. However, in the third model (C) the network recombines only to the point of $\langle S\rangle \approx 0.2N$ and nodes with no edges still have a chance to gain a connection, but their probability is negligible compared to the probability for hubs. The fourth model (D) recombines to $\langle S\rangle \approx 0.2N$. Again, the lonely nodes cannot be chosen for neighbors and switching is a local event in this case. This model displays quite a different behavior for small $q$ - its phase diagram is convex and it has the smallest size of the phase I (only $q = 2$, see FIG.~\ref{fig:phases}).

\begin{figure}
\centering
  \includegraphics[width=0.75\linewidth]{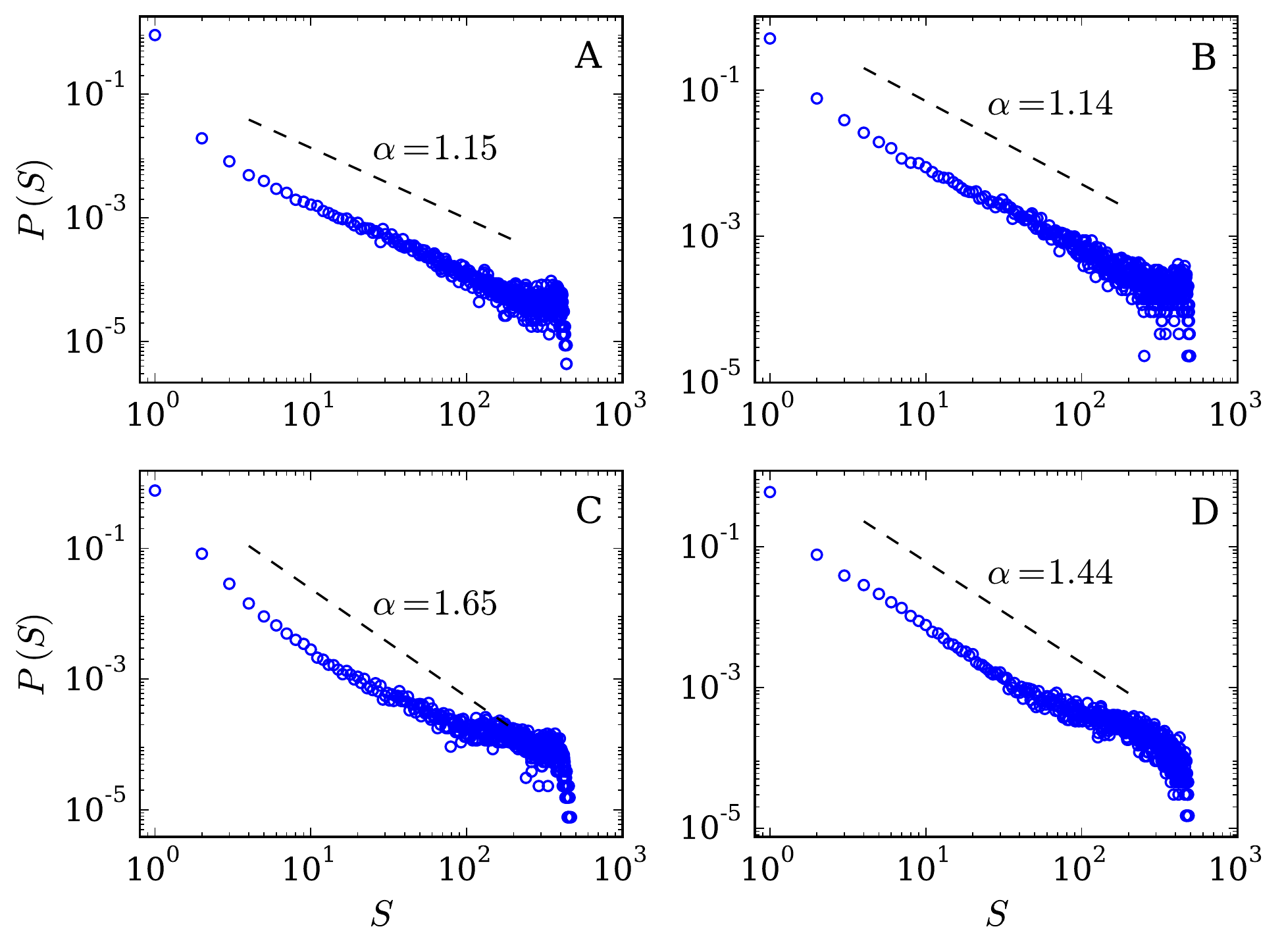}
\caption{Size distribution of network components at the first transition point $q_c$. Dashed lines represent best fit of the power law.
Character in the upper right corner represents the~number of a model.\label{fig:components}}
\end{figure}

We identify both transition points $q_c$ and $q^*$, firstly, by an abrupt change of network's topology resulting in a sudden increase or decrease of the order parameter $\langle S\rangle$ and secondly, because both of them are characterized by a maximum of fluctuations. Additionally, unlike for any other value of $q$, in the first transition point $q_c$ we obtain a power-law distribution of the size of the components (FIG. \ref{fig:components}). This property is universal for all models - even the model D. We have also discovered that the second transition point $q^*$ shows yet another special criticality characteristic, again universal for all models (not only those with long-distance switching), namely a significant increase of the clustering coefficient (see FIG. \ref{fig:phases}). At this point the domains contain at most several nodes, and the number of edges and pairs of compatible agents (nodes with one or more common traits) is the same. In consequence, every pair of nodes within every domain will eventually be connected, because the nodes of one domain are compatible agents. This leads to a formation of numerous complete subgraphs in the network. In effect, the clustering coefficient takes the maximal value at the point $q^*$. This observation is valid even in the model D, although in this case the clustering coefficient is high for other values of $q$ as well, due to the switching mechanism. The mean-field equation for the second transition point $q^*=NF/\langle k\rangle$ \cite{sanmiguel2007} is no longer valid in our models.

\begin{figure}
\centering
  \includegraphics[width=0.75\linewidth]{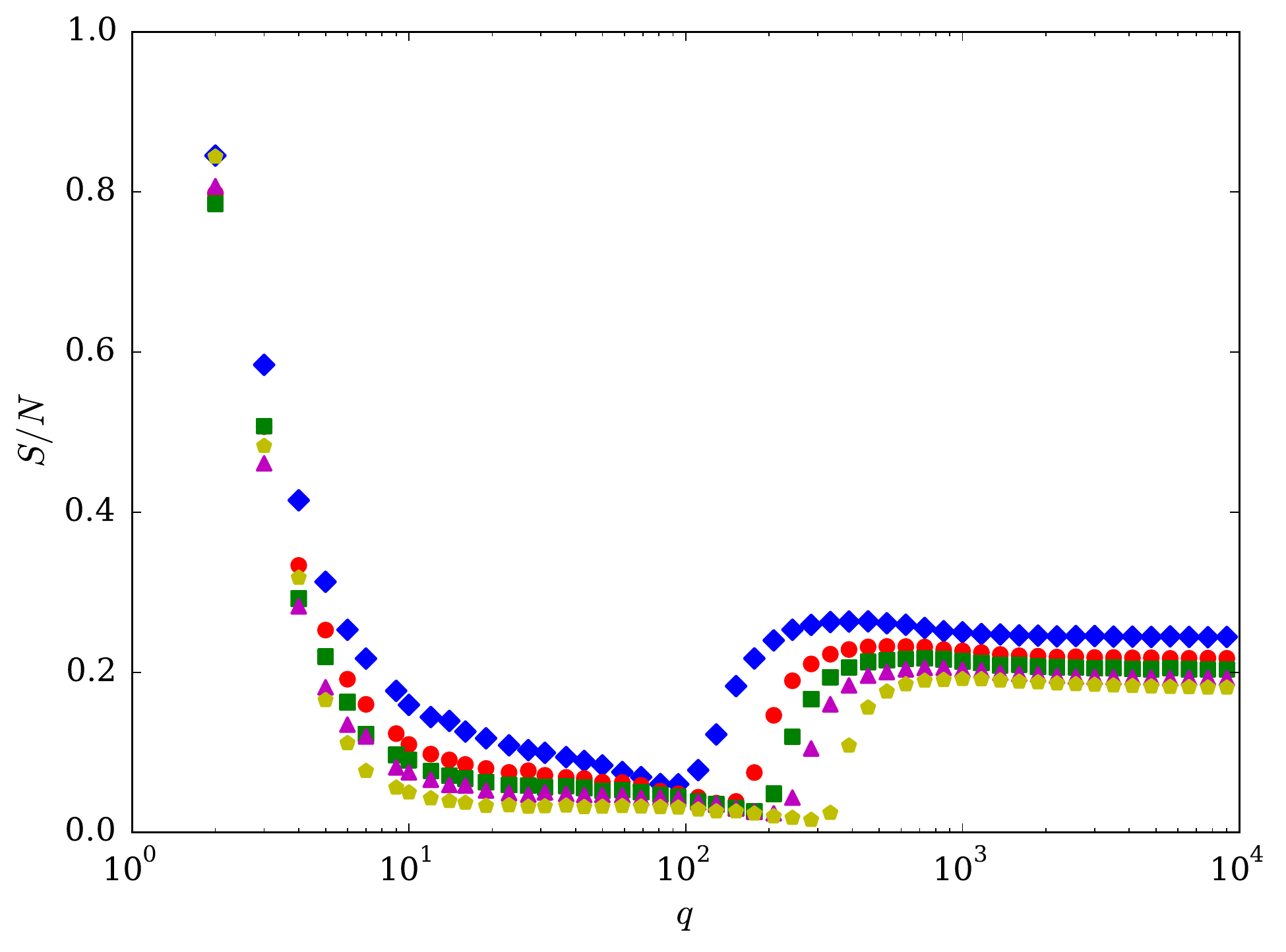}
\caption{Average relative size of the largest network component for the model D, for $N = 500$ (diamonds), $N = 1000$ (circles), $N = 1500$ (squares), $N = 2000$ (triangles) and $N = 4000$ (pentagons), averaged over 400 realizations. \label{fig:scaling}}
\end{figure}

However, the model of nearest neighborhood (D) displays also a different scaling behavior. We have run simulations for several sizes $N$ of the network. In the first three models (A, B, and C), just as in the original model, the transition point $q_c$ increases with the system size $N$, suggesting that the transition becomes continuous in the thermodynamic limit. In the model D the transition remains at $q_c = 3$ regardless of $N$, suggesting discontinuous order-disorder transition in the thermodynamic limit (see FIG.~\ref{fig:scaling}). At the same time, the second transition point $q^*$ increases with the system size $N$ in all four cases.

%%%%%%%%%%%%%%%%%%%%%%%%%%%%%%%%%
%                                                                                       %
%                      STRUCTERE OF NETWORKS                        %
%                                                                                       %
%%%%%%%%%%%%%%%%%%%%%%%%%%%%%%%%%

\section{Structure of networks}\label{section:stucture}

\begin{figure}
\centering
  \includegraphics[width=1.0\linewidth]{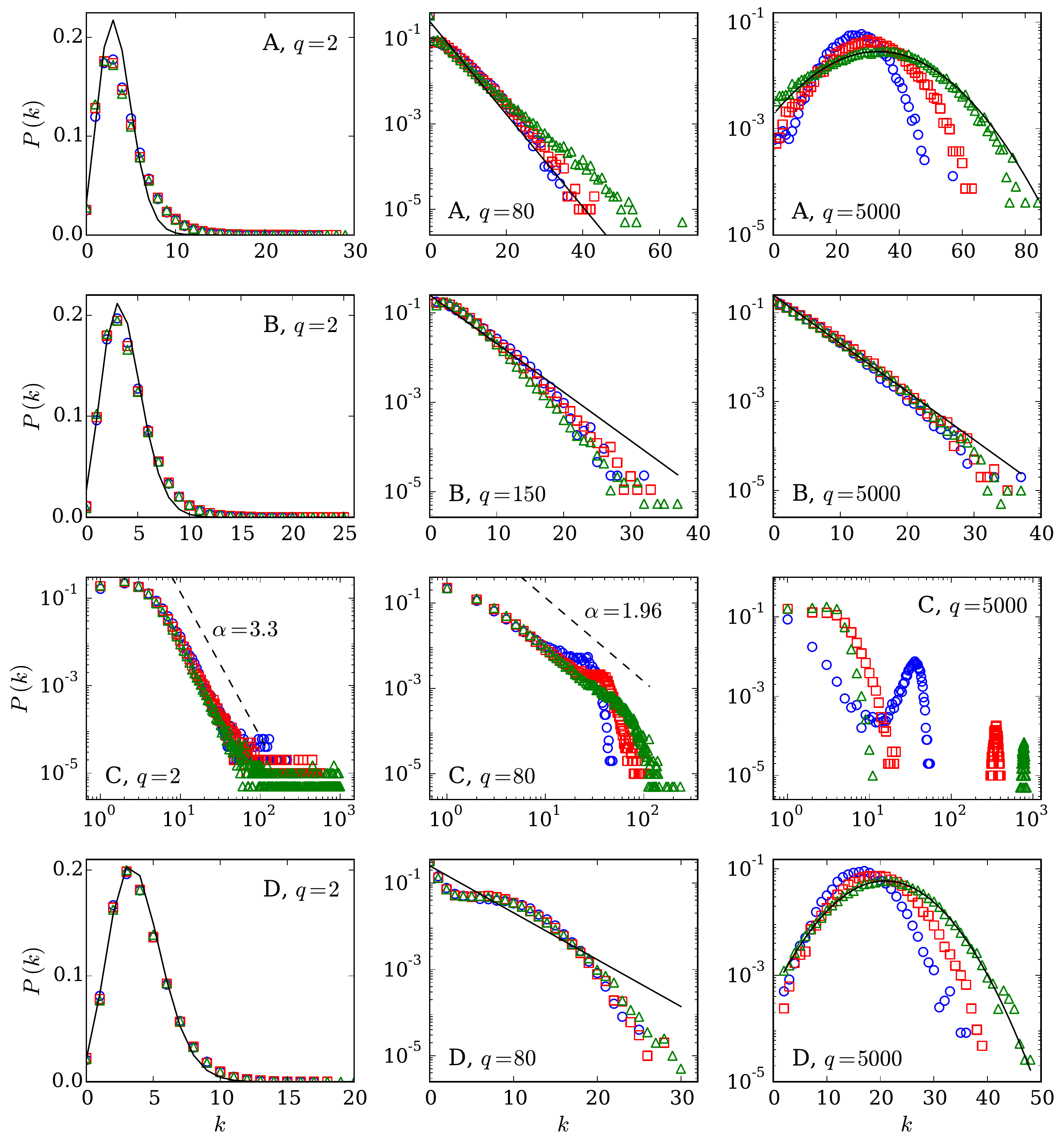}
\caption{Degree distribution in the stationary state for $N = 500$ (circles), $N = 1000$ (squares) and $N = 2000$ (triangles), averaged over 100 realizations.
Dashed lines represent the power law, solid lines represent exponential, Gaussian or Poisson distribution.
The number of a~model and the value of parameter $q$ are given in the panel. \label{fig:degree}}
\end{figure}

In order to comprehensively describe the structure of the networks generated by presented models in all phases, we examined the degree distribution, the average path length and the clustering coefficient\footnote{We have calculated both global and local clustering coefficients. When a given node has less than two neighbors its   local clustering coefficient cannot be calculated and it is, therefore, not taken into account.}. The resulting structure is more complex, than in the original model. The degree distribution is not always Poisson and varies between models and between phases (see FIG. \ref{fig:degree}).

We considered four possible distributions: Poisson, Gaussian, exponential and power-law. For every model in every phase, all four distributions were fitted using the least-squares method, with one exception - if the empirical distribution was strongly asymetric or monotonically decreasing, we ruled out Gaussian distribution, as it has different natrue.
Subsequently, we calculated the coefficient of determination $R^2$ and we compared its value among fits. Distributions with the value closest to $1$ were chosen as the best approximation of the empirical data. If, for a given model in a given phase, the coefficient $R^2 < 0.9$ for all fits or the empirical distribution was discontinuous we marked it as unclassified.
 In the results listed below,
by Poisson / Exponential we mean, that the tail of the distribution is bigger than Poisson and smaller than exponential, but in both cases Poisson distribution is a better fit in terms of $R^2$ value. In the phase II and III, we omit nodes with no connections, since they bias the resulting distribution.

\begin{enumerate}
\item The model with preference of high degree (A):
	\begin{itemize}
    \item Phase I: Poisson / Exponential
    \item Phase II: Exponential
    \item Phase III: Gaussian
    \end{itemize}
\item The model with preference of high degree with auto-connections (B):
	\begin{itemize}
    \item Phase I: Poisson / Exponential
    \item Phase II: Exponential
    \item Phase III: Exponential
    \end{itemize}
\item The model with square preference of high degree (C):
	\begin{itemize}
    \item Phase I: Power-law
    \item Phase II: Power-law
    \item Phase III: Unclassified
    \end{itemize}
\item The model with preference of close neighborhood (D):
	\begin{itemize}
    \item Phase I: Poisson
    \item Phase II: Unclassified
    \item Phase III: Gaussian
    \end{itemize}
\end{enumerate}

For every phase and every model a plot of the degree distribution is shown in FIG. \ref{fig:degree}. In the model A, in the phase II the distribution is approximately exponential, but as observed for bigger networks tail is getting fatter. No such behavior occurs in the second model (B) - the distributions have exponential tails in the phase II regardless of the network size. The most interesting results, however, can be found in the third model. In the first two phases the degree distribution is power-law and the data collapses for all sizes of the network. Furthermore, for bigger networks, the distribution matches the power function in a wider scope, indicating that the bend at the end of the distribution is a finite-size effect. In the phase III the distribution takes an unusual form, and is discontinuous for large $N$. This form results in an approximately power-law decay for small degree and a narrow peak for large degree. The fourth model (D) has Poisson distribution in the phase I, in the phase II it is unclassified, but has approximately exponential tail, and in the third phase it takes Gaussian form.

Despite the fact that the initial network was always poissonian, even after a relatively small number of switches, i.e. for $q \approx 2$, the graphs generated in our models are different from those in the original model and different from the canonical Erd{\H{o}}s-R{\'e}nyi random graphs. First of all, in every presented model the maximal degree is always greater. Secondly, the average path length can be different (see below). Finally, the clustering coefficient is higher. For instance, it takes values from $0.4$ to $0.8$ in the phase I and II for the model D and $0.35$ in the phase III for the model A (see FIG.~\ref{fig:phases}), while in the original model the clustering coefficient is around $0.02$ (except for the second transition point), and in Erd{\H{o}}s-R{\'e}nyi graph it is around $0.01$ for the same system size and average degree \cite{newman2002random}.

\begin{figure}
\centering
  \includegraphics[width=0.75\linewidth]{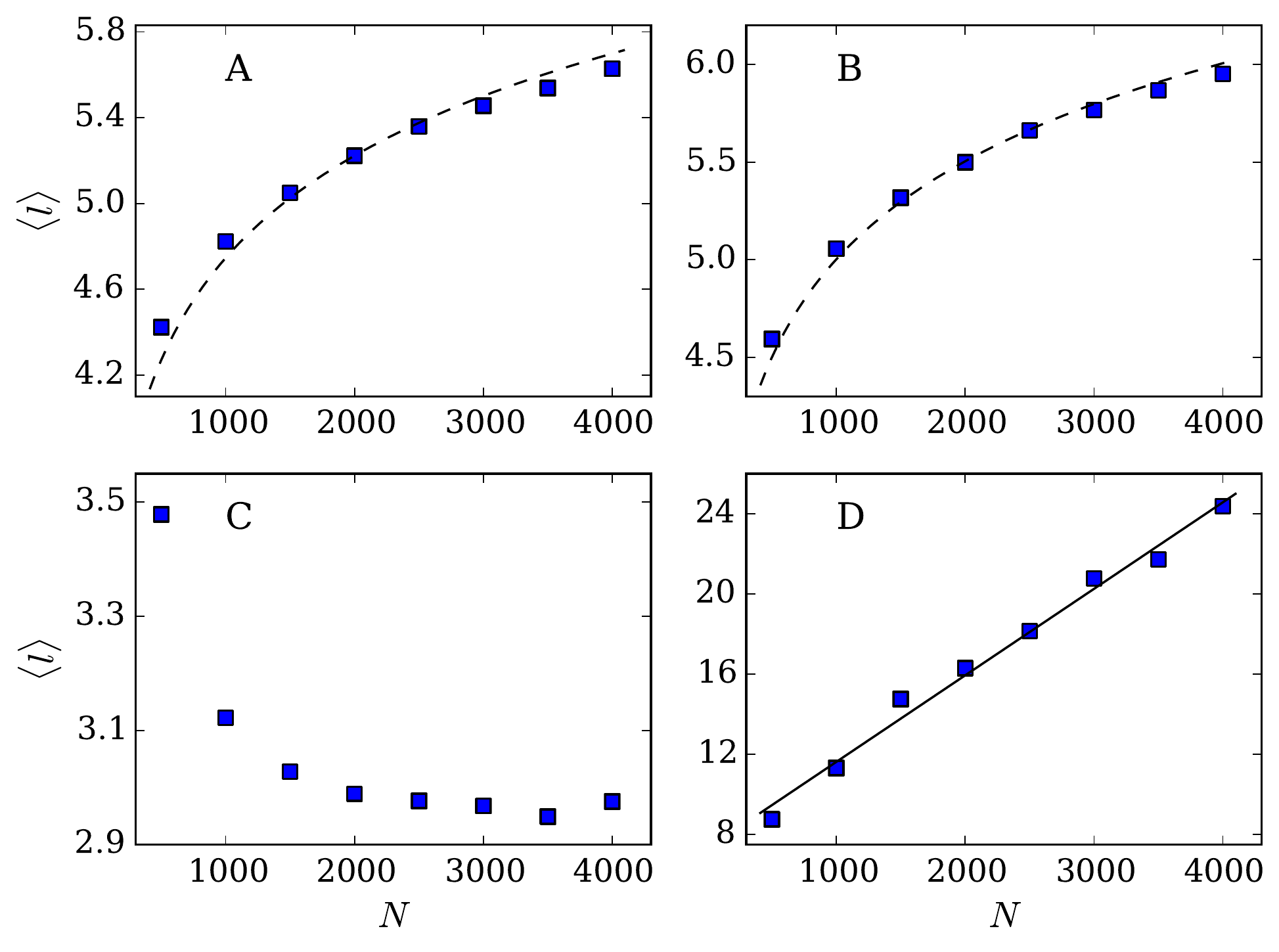}
\caption{Average path length $\langle l \rangle$ vs system size $N$. Dashed lines represent a logarithmic expansion, solid line represents a linear function.
Character in the upper left corner represents the~number of a model.\label{fig:path}}
\end{figure}

We analyzed the average path length $\langle l \rangle$ for all models (FIG. \ref{fig:path}). Models A, B, and C generate networks with the small-world effect \cite{watts1998collective,milgram1967small,travers1969experimental}. In the models with the preference of high degree (A and B), the average path length $\langle l \rangle$ grows with size $N$ of the network logarithmically, or even a bit slower, due to long-distant switching. In the model with the square preference of high degree (C), the average path length initially decreases with growth of the network to stabilize around~$3$. This is a result of large hubs formation in bigger networks, with a degree comparable to the size of the network, yet smaller than\footnote{A network containing a node with a maximal degree equal $N-1$, would have average path length smaller than $2$.} $N-1$. The model with the preference of close neighbors (D) does not generate networks with small-world effect. The average path length grows linearly with size of the network. The initial graph is always random, and thus displays small-world property, but the local rewiring mechanism erases all long-distance connections and replaces them with short connections in nearest neighborhood. We have calculated the average path length only for the first phase to avoid definition issues when a graph is not connected.

%%%%%%%%%%%%%%%%%%%%%%%%%%%%%%%%%
%                                                                                       %
%                                  SUMMARY                                      %
%                                                                                       %
%%%%%%%%%%%%%%%%%%%%%%%%%%%%%%%%%

\section{Summary}\label{section:summary}

We studied four models of social interactions with coevolution of states of nodes and dynamics of network's topology. All models exhibit three phases - the static phase~I with an ordered final state, the static phase II with a disordered final state, and the dynamic phase III, where a recombination occurs. Both phase transition points  are described by an abrupt change of network's topology and a maximum of fluctuations. In addition, the first transition point $q_c$ is characterized by the power-law distribution of component's size, and the second point $q^*$ is characterized by maximal value of the clustering coefficient. In the first three models (A, B, C) both $q_c$ and $q^*$ increase with the system size $N$, while in the model D only the latter transition point ($q^*$) increases, while the former ($q_c$) remains at $q=3$.

Another new result is a wide variety of generated networks. Depending on the model and the phase we can observe Poisson, exponential, Gaussian, or power-law degree distribution and different types of mixed distributions. In the model C in the phase I and II we obtain the power-law distribution with the same exponent for different system sizes, stretching from $\alpha=3.5$ in the phase~I to $\alpha=1.9$ in the phase~II. In graphs generated using the models A and B the small-world effect is present in the sense of a logarithmic growth of the average path length with the system size. The model C displays an even stronger effect - the average path length decreases with the system size to stabilize around $3$. In the last model (D) the small-world effect is absent and the average path length grows linearly.

We have shown that the procedure of creating new links in coevolving networks can significantly affect the structure of the system, but analytical solutions are still desirable. The signature sign of the second transition point we discovered - the peak of the clustering coefficient - is especially interesting in the context of social groups, where clustering is very common, as it has been shown by multiple studies. Our results provide a simple method, which could explain different structures and dynamics in coevolving complex networks, e. g. social systems.

%%%%%%%%%%%%%%%%%%%%%%%%%%%%%%%%%
%                                                                                       %
%                 ACKNOWLEDGMENTS AND APPENDIX               %
%                                                                                       %
%%%%%%%%%%%%%%%%%%%%%%%%%%%%%%%%%

\section*{Acknowledgments}\label{section:acknowledgments}
The authors would like to thank Ryszard Kutner for fruitful discussions and Piotr Ochnicki for patient inspections.

\pagebreak

\appendix
\section{Comparison of the results in all models}\label{section:appendix}

\begin{figure}[h]
\captionsetup[subfigure]{labelformat=empty}
\begin{subfigure}{.5\textwidth}
  \centering
  \includegraphics[width=1.0\linewidth]{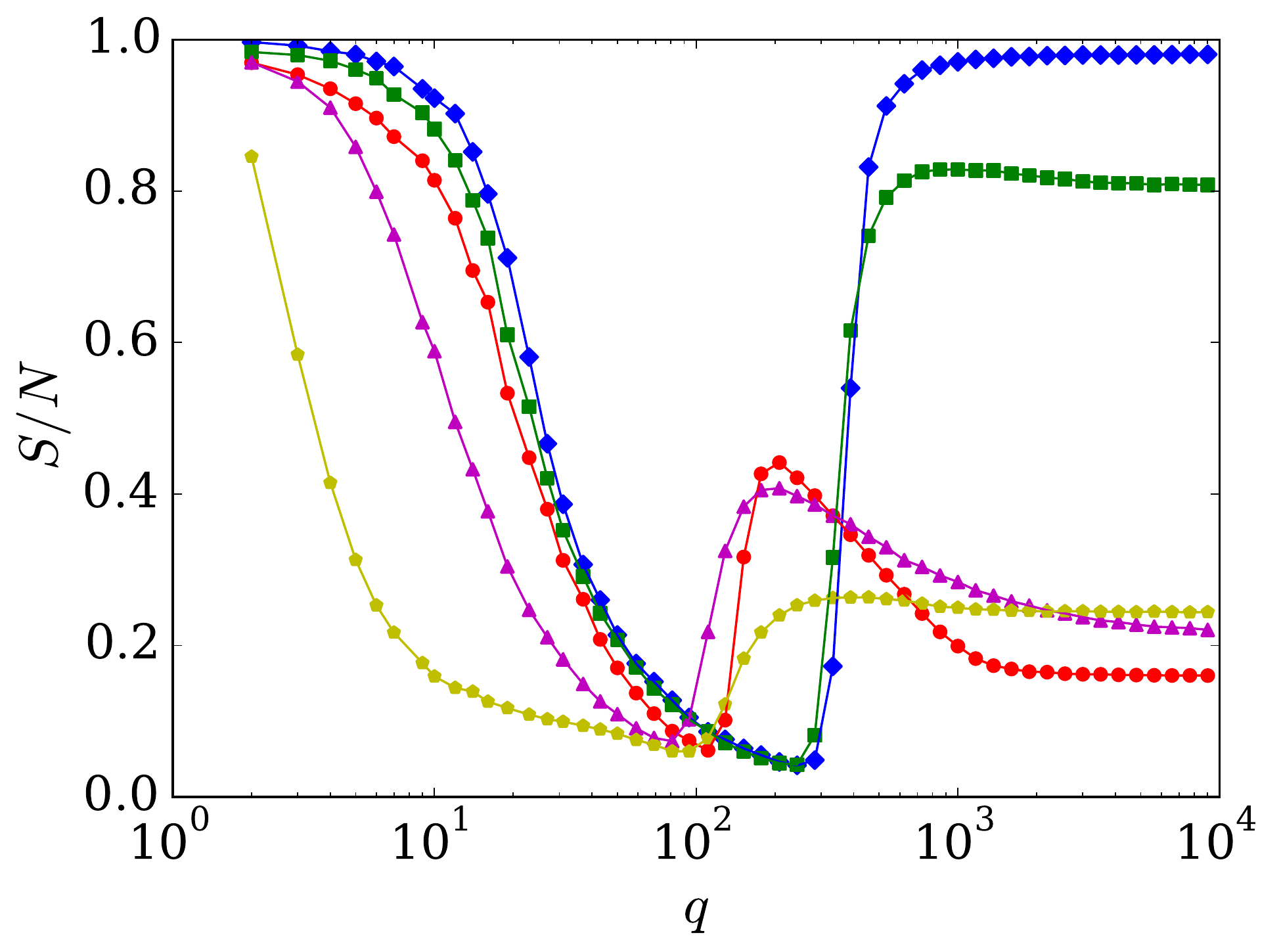}
  \caption{\label{sub:BA}}
\end{subfigure}
\begin{subfigure}{.5\textwidth}
  \centering
  \includegraphics[width=1.0\linewidth]{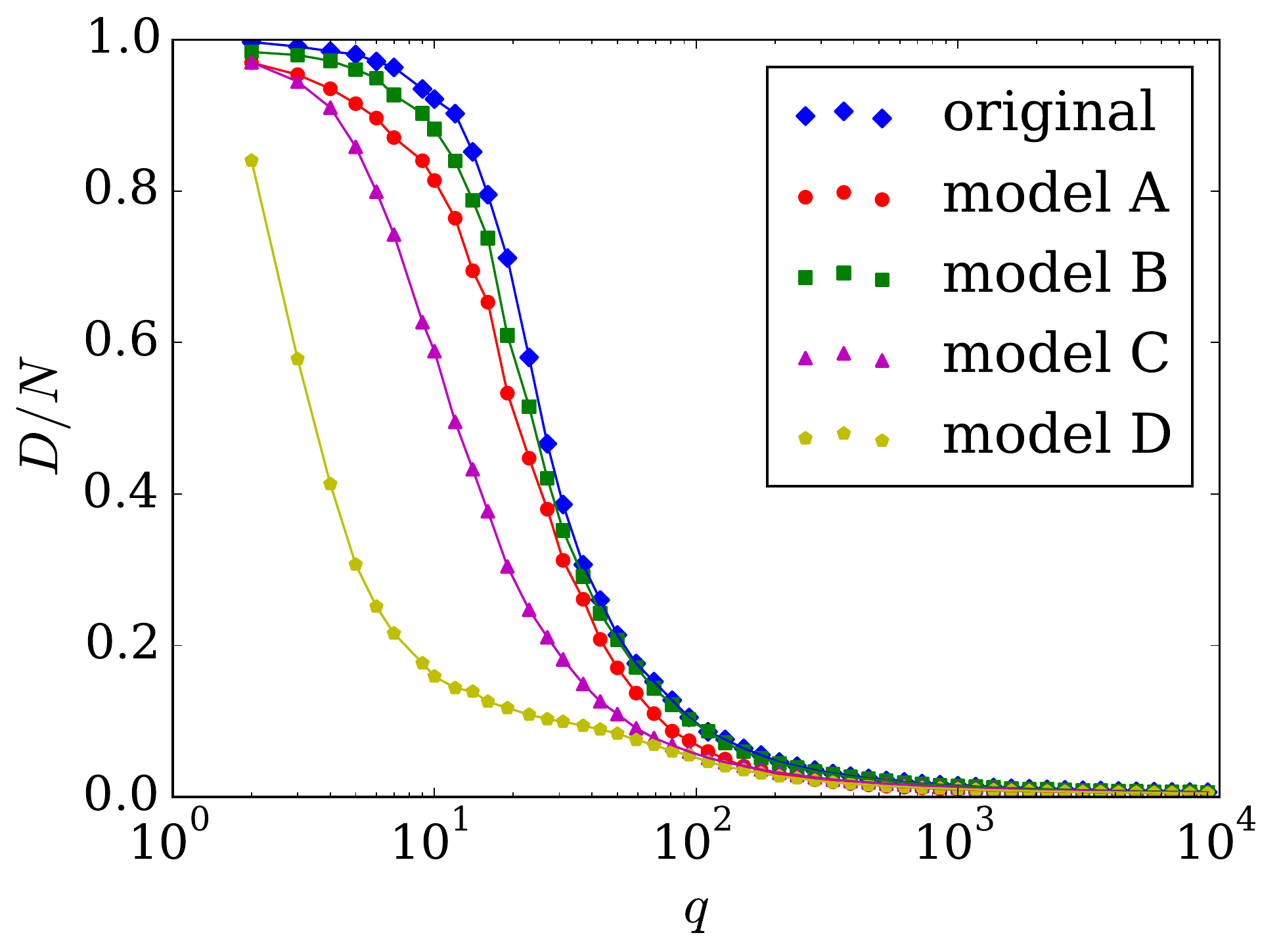}
  \caption{\label{sub:k1}}
\end{subfigure}
\begin{subfigure}{.5\textwidth}
  \centering
  \includegraphics[width=1.0\linewidth]{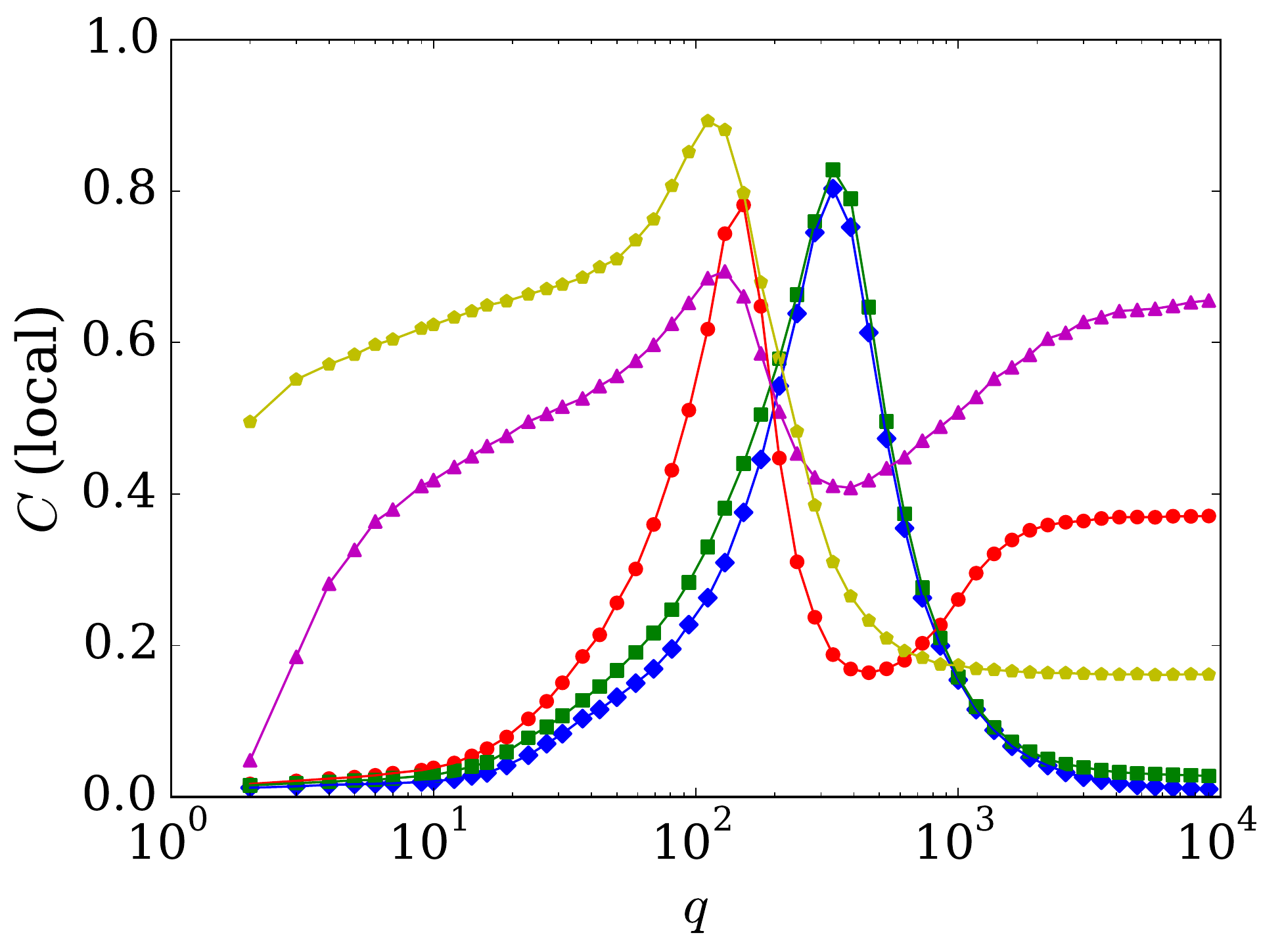}
  \caption{\label{sub:k2}}
\end{subfigure}
\begin{subfigure}{.5\textwidth}
  \centering
  \includegraphics[width=1.0\linewidth]{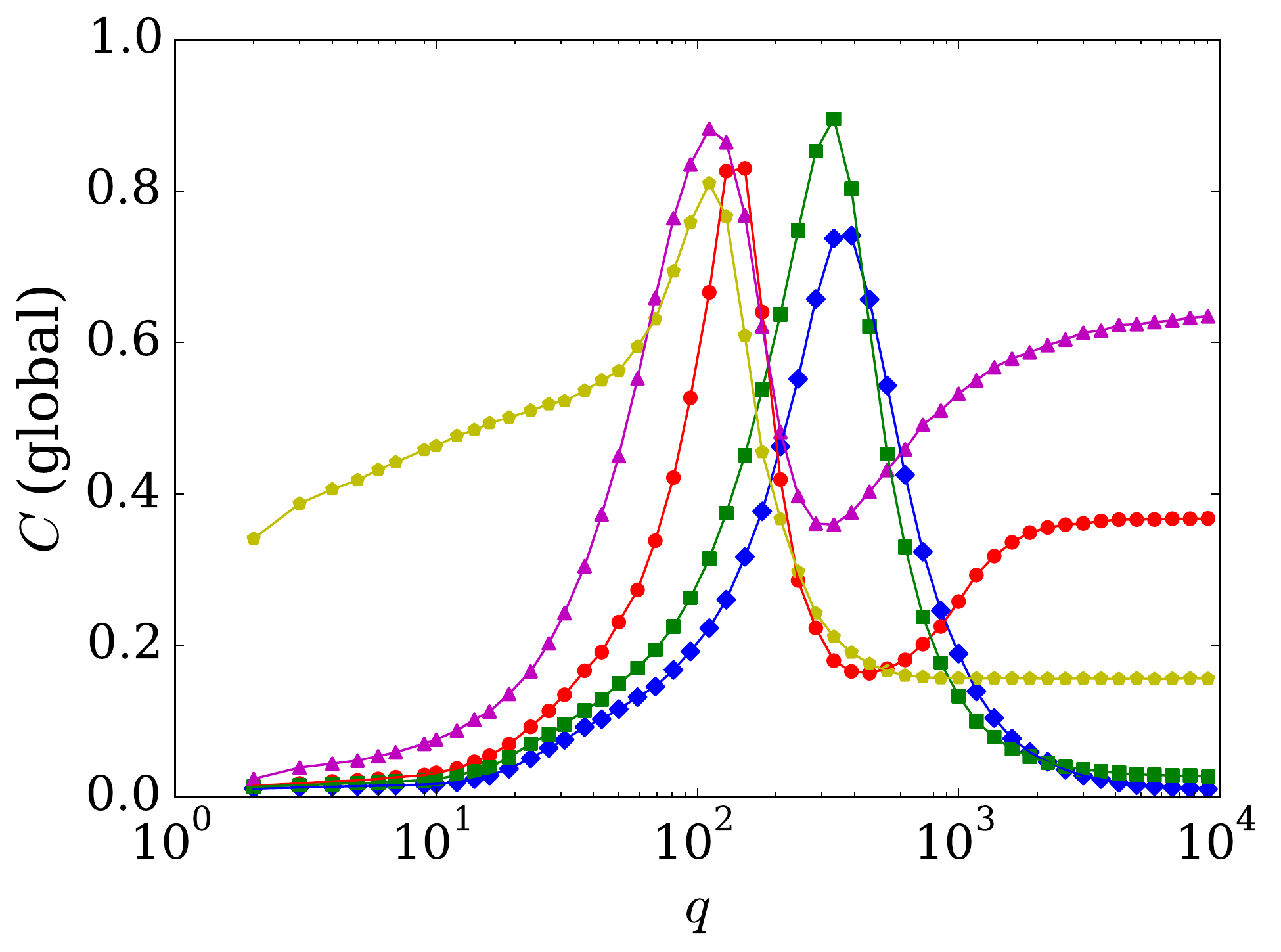}
  \caption{\label{sub:clust}}
\end{subfigure}
\caption{Comparison of the average relative size of the largest network component (upper left) and largest domain (upper right),
global (lower left) and average local (lower right) clustering coefficient in the stationary configuration vs $q$, for $N = 500$, averaged over 400 realizations.
Blue diamonds - original model, red circles - model A, green squares - model B, purple triangles - model C, yellow pentagons - model D
(as stated in the legend).\label{fig:phases2}}
\end{figure}

%%%%%%%%%%%%%%%%%%%%%%%%%%%%%%%%%
%                                                                                       %
%                             BIBLIOGRAPHY                                     %
%                                                                                       %
%%%%%%%%%%%%%%%%%%%%%%%%%%%%%%%%%

%\nocite{*}
\bibliographystyle{ieeetr}
\bibliography{bibliografia}

\end{document}